# X-Ray Diffraction Studies of Copper Nanopowder

**T. Theivasanthi** [(1)] **and M. Alagar** [(2)]
(1) Department of Physics, PACR Polytechnic College, Rajapalayam, India.
(2) Department of Physics, Ayya Nadar Janaki Ammal College, Sivakasi, India.

**Abstract:** *Copper nanopowder preparation and its X-Ray diffraction studies are reported in this paper. Electrolytic cathode deposition method is simple and cheapest process for its preparation. Copper nanopowder has been prepared from aqueous copper sulphate solution. Wide range of experimental conditions has been adopted in this process and its X-Ray diffraction characterizations have been studied. The results confirming copper nanopowder with size below 30 nm. Uniformed size Copper nanopowder preparation, in normal room temperature is importance of this study.*

## 1. Introduction

Nanoparticle preparation and study about nanoparticle are importance in the recent research [1-3]. The characters of metal nano particles like optical, electronic, magnetic, and catalytic are depending on their size, shape and chemical surroundings [2, 3]. In nanoparticle preparation it is very important to control the particle size, particle shape and morphology. XRD study is most important tool used in nano materials science. A discussion about simple and low cost preparation of Copper nanopowder and its X-ray diffractional (XRD) studies are presented in this study. Preparation of uniformed Copper nanopowder size less than 30 nm, in a normal room temperature is importance of this study. Its XRD analysis confirms the result.

Nanosized metal particles are attracting the attention of present science field because of their physical and chemical properties, which are quite dissimilar from those of bulk materials.[4]. Various techniques have been adopted to produce nanoparticles on solid surfaces, including diverse lithographic techniques, vacuum deposition of metal, controlled nanoparticle growth by diffusion, electrophoretic deposition of a metal colloid, chemical and electrochemical deposition of metal nanoparticles, etc.[12-15]. Nanoparticle synthesized in Lithographic and vacuum deposition of metal techniques uniform in size, shape, and spacing but these techniques are expensive, suitable for small number of materials systems. Electrolytic cathode deposition is one of the suitable, simplest and low-cost methods and can be used in wide range of materials.

## 2. Copper Nanopowder Preparation

Electrolytic cathode deposition method was adopted for Copper nanopowder preparation. Copper Sulpahte was kept in a cleaned glass vessel, water was poured and a homogenous aqueous copper sulphate solution was made. Surface-cleaned anode and cathode copper plates were kept inside the glass vessel. Electrolysis of this solution was done by passing 1.5 A current inside solution through anode and cathode. A wide range of experimental conditions has been adopted with constant current (1.5 A) and with constant concentration of the Copper Sulpahte solution for various time periods, in a normal room temperature. At the end of electrolyzing process, a layer of copper deposition on the cathode surface was observed. Copper layer was removed from the cathode surface and got copper nanopowder. An inorganic confirmation test for copper was conducted. Its structural characterizations were studied by using X-ray diffraction. The results confirm the formation of copper nanopowder of diameter less than 30 nm.

_________________________________________________
[(1)] **Corresponding author.**     *E-mail*:  sankarg4@yahoo.com

### 2.1. Reactions of Copper Nano powder

Inorganic qualitative test was conducted for confirmation of copper. A small quantity of prepared powder was taken in a test tube and dilute sodium hydroxide solution was added with stirring. A Pale blue colour precipitation was observed in the test tube [9]. The test result confirms the prepared material is Copper. The result is presented in Fig.1

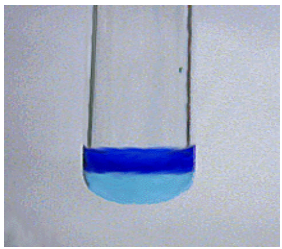

Fig. 1

## 3. X-Ray Diffraction Studies

Powder X-ray Diffraction (XRD) is one of the primary techniques used by mineralogists and solid state chemists to examine the physico-chemical make-up of unknown materials.

X-ray diffraction is one of the most important characterization tools used in solid state chemistry and materials science. XRD is an easy tool to determine the size and the shape of the unit cell for any compound. Powder Diffraction Methods is useful for Qualitative analysis (Phase Identification), Quantitative analysis (Lattice parameter determination & Phase fraction analysis) etc. *Diffraction pattern* gives information on translational symmetry - size and shape of the unit cell from *Peak Positions* and information on electron density inside the unit cell, namely where the atoms are located from *Peak Intensities*. It also gives information on deviations from a perfect particle, if size is less than roughly 100 – 200nm, extended defects and micro strain from *Peak Shapes & Widths*

### 3.1. Peak Indexing

Indexing is the process of determining the unit cell dimensions from the peak positions. It is the first step in diffraction pattern analysis. To index a powder diffraction pattern it is necessary to assign *Miller Indices* (h k l) to each peak. Unfortunately it is not just the simple reverse of calculating peak positions from the unit cell dimensions and wavelength [16].

Table.1: XRD Data

| $2\theta^0$ | Counts |
|---|---|
| 10 | 36.6667 |
| 10.02 | 33.3333 |
| 10.04 | 56.6667 |
| 10.06 | 20 |
| 10.08 | 26.6667 |
| 10.1 | 36.6667 |
| ....Skip some lines.... | |
| 79.9 | 163.333 |
| 79.92 | 156.667 |
| 79.94 | 126.667 |
| 79.96 | 186.667 |
| 79.98 | 193.333 |
| 80 | 170 |

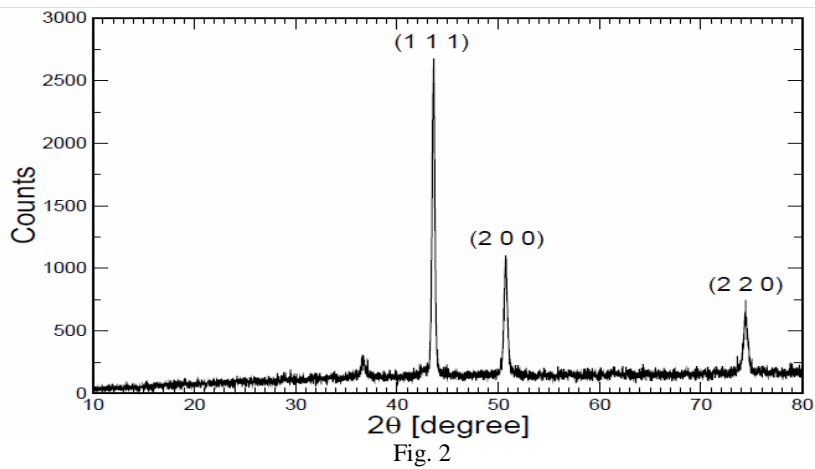

Fig. 2

XRD analysis of the prepared sample of Copper nanopowder was done by a Goniometer (Ultima3 theta-theta gonio, under 40kV/30mA - X-Ray, 2θ / θ – Scanning mode, Fixed Monochromator). Data was taken for the 2θ range of 10 to 80 degrees with a step of 0.02 degree. Data for some 2θ range has given in Table.1. Indexing process of powder diffraction pattern was done and *Miller Indices* (h k l) to each peak was assigned in first step. Diffractogram of the entire data is in Fig.2.

Table.2: Simple peak indexing.

| Peak position, 2θ | 1000×Sin²θ | 1000×Sin²θ / 46 | Reflection | Remarks |
|---|---|---|---|---|
| 43.6 | 138 | 3 | (1 1 1) | $1^2+1^2+1^2 = 3$ |
| 50.8 | 184 | 4 | (2 0 0) | $2^2+0^2+0^2 = 4$ |
| 74.4 | 366 | 8 | (2 2 0) | $2^2+2^2+0^2 = 8$ |

Indexing has been done in two different methods and data are in Table.2 & Table.3. In table.2, one need to find a dividing constant and the values in the 3rd column becomes integers (approximately). Here, the constant is 46(= 184−138). Moreover, the high intense peak for FCC materials is generally (1 1 1) reflection, which is observed in the sample.

Table.3: Peak indexing from d – spacing

| 2θ | d | 1000/d² | (1000/d²) / 77.32 | hkl |
|---|---|---|---|---|
| 43.64 | 2.073 | 232.07 | 3.01 | 111 |
| 50.80 | 1.796 | 310.02 | 4.01 | 200 |
| 74.42 | 1.274 | 616.11 | 8.00 | 220 |

Table.4: Experimental and standard diffraction angles of Cu specimen

| Experimental diffraction angle [2θ in degrees] | Standard diffraction angle [2θ in degrees] JCPDS Copper: 04-0836 |
|---|---|
| 43.640 | 43.297 |
| 50.800 | 50.433 |
| 74.420 | 74.130 |

Three peaks at 2θ values of 43.640, 50.800, and 74.420 deg corresponding to (111), (200), and (220) planes of copper were observed and compared with the standard powder diffraction card of JCPDS, copper file No. 04–0836. Table.4 shows the experimentally obtained X-ray diffraction angle and the standard diffraction angle of Cu specimen. The XRD study confirms / indicates that the resultant particles are (FCC) Copper Nanopowder.

### 3.2. Particle Size Calculation

From this study, considering the peak at degrees, average particle size has been estimated by using *Debye-Scherrer formula* [5-8]

$$D = \frac{0.9\,\lambda}{\beta \cos \theta} \quad \ldots\ldots\ldots\ldots\ldots\ldots\ldots (1)$$

Where 'λ' is wave length of X-Ray (0.1541 nm), 'β' is FWHM (full width at half maximum),
'θ' is the diffraction angle and 'D' is particle diameter size.

1) $2\theta = 43.64$
   $\beta = (43.80 - 43.46) \times 3.14 / 180 = 0.0059$ radians
   $$D = \frac{0.9 \times 0.1541}{0.0059 \times \cos 21.82} = 25 \text{ nm}$$

2) $2\theta = 50.80$
   $\beta = (50.92 - 50.54) \times 3.14 / 180 = 0.0066$ radians
   $$D = \frac{0.9 \times 0.1541}{0.0066 \times \cos 25.4} = 23.26 \text{ nm}$$

3) $2\theta = 74.42$
   $\beta = (74.68 - 74.26) \times 3.14 / 180 = 0.0024$ radians
   $$D = \frac{0.9 \times 0.1541}{0.0024 \times \cos 37.21} = 24.88 \text{ nm}$$

The particle size is less than 30 nm and the details are in Table.5.

### 3.1. Calculation of d-Spacing

The value of d (the interplanar spacing between the atoms) is calculated using

*Bragg's Law:* $2d\sin\theta = n\lambda$ ………………….. (2)

$$d = \frac{\lambda}{2\sin\theta} \quad (n=1)$$

Wavelength $\lambda = 1.5418$ Å for Cu Ka

1) $2\theta = 43.64$
   $\theta = 21.82$
   $$d = \frac{0.1541}{2\sin 21.82} = 0.2073 \text{ nm}$$

2) $2\theta = 50.80$
   $\theta = 25.40$
   $$d = \frac{0.1541}{2\sin 25.40} = 0.1796 \text{ nm}$$

3) $2\theta = 74.42$
   $\theta = 37.21$
   $$d = \frac{0.1541}{2\sin 37.21} = 0.1274 \text{ nm}$$

The calculated d-spacing details are in Table.5.

### 3.2. Calculation for expected 2θ positions of the first three peaks

The crystal structure of copper is face centered cubic, with a unit cell edge a = 3.60 Å
(a=4/√2 x r. r =128pm for copper). Following formulas are used in the calculation of the *expected 2θ positions of the first three peaks* in the diffraction pattern and the *interplanar spacing d* for each peak.

$$1/d^2 = (h^2+k^2+l^2) / a^2 \quad\ldots\ldots\ldots\ldots\ldots\ldots\ldots. \quad (3)$$

*Bragg's Law* is used to determine the 2θ value: $\lambda = 2d_{hkl} \sin \theta_{hkl}$

1) hkl = 111
$1/d^2 = (1^2 + 1^2 + 1^2) / (3.60 \text{ Å})^2 \rightarrow d = 2.078$ Å
$\sin \theta_{111} = 1.54$ Å$/\{2(2.078$ Å$)\} \rightarrow \theta = 21.75° (2\theta = 43.5°)$

2) hkl = 200
$1/d^2 = (2^2 + 0^2 + 0^2) / (3.60 \text{ Å})^2 \rightarrow d = 1.8$ Å
$\sin \theta_{200} = 1.54$ Å$/\{2(1.8$ Å$)\} \rightarrow \theta = 25.3° (2\theta = 50.6°)$

3) hkl = 220
$1/d^2 = (2^2 + 2^2 + 0^2) / (3.60 \text{ Å})^2 \rightarrow d = 1.274$ Å
$\sin \theta_{220} = 1.54$ Å$/\{2(1.274$ Å$)\} \rightarrow \theta = 37.25° (2\theta = 74.5°)$

These expected 2θ values are very close with the experimental 2θ values mentioned in the Table.5.

Table.5: The grain size of Copper nanopowder

| 2θ of the intense peak (deg) | hkl | θ of the intense peak (deg) | FWHM of Intense peak (β) radians | Size of the partcle (D) nm | d-spacing nm |
|---|---|---|---|---|---|
| 43.64 | (111) | 21.82 | 0.0059 | 25.32 | 0.2073 |
| 50.80 | (200) | 25.40 | 0.0066 | 23.26 | 0.1796 |
| 74.42 | (220) | 37.21 | 0.0070 | 24.88 | 0.1274 |

## 4. Results and Discussions

Copper nanoparticles have already been studied extensively due to their potential technological applications in various fields like catalysis, lubricants, electronics etc.[11]. X-ray diffraction is an easy and one of the most important characterization tools used in nanomaterial research field. Here, an important nanomaterial - *copper nanopowder* has been successfully prepared in *electrolytic cathode deposition method* in normal room temperature and its structural characterizations have been studied by important tool - *X-ray diffraction.* The results confirm nanopowder, uniformed size less than 30nm.

Three peaks at 2θ values of 43.64, 50.80, and 74.42 deg corresponding to (111), (200), and (220) planes of copper have been observed and compared with the JCPDS, copper file No. 04–0836 and ASTM 03-1005- face-centered cubic copper phase - standard powder diffraction card. [10]. The said 2θ values of three peaks are in accordance with the standard of both JCPDS & ASTM. The XRD study confirms / indicates that the resultant particles are (FCC) Copper Nanopowder.

In addition to this, the experimental powder diffraction pattern presented in Fig.3 have been compared with Fig.4 Copper powder pattern like fingerprint verification. The diffraction pattern in Fig. 4 is a Database / Picture of *NT Base Co., Ltd.* (Seoul, 449-934, Korea) and it has been retrieved from the *Nanomaterial Database™* of *Nanowerk LLC*, Honolulu, HI 96813, USA. Both the powder diffraction patterns are in matching condition.

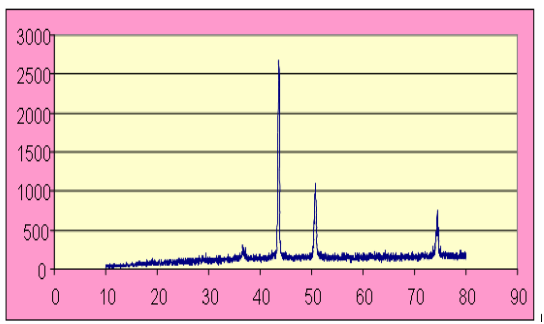 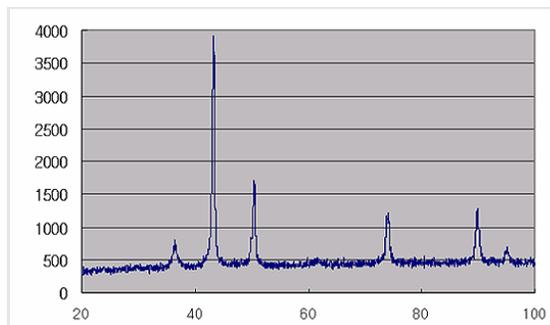

Fig. 3                                                                Fig. 4

## 5. Conclusion

A new method, electrolytic cathode deposition has been applied in Copper Nanopowder preparation and successfully completed the entire process. Based on this study, some other nanopowder may be prepared in future.

## 6. Acknowledgements

The author expresses immense thanks to staff and management of PACR Polytechnic College, Rajapalayam, India, Mr.S.Rajagopalan, Institut fuer Physikalische Chemie, Stuttgart University, Germany and R. Srinivasan, Dept. of Physics, National Institute of Technology, Tiruchirappalli, India for valuable suggestions, assistance and encouragement during this work.